\documentclass[lettersize,journal]{IEEEtran}

\usepackage[nocompress]{cite}
\usepackage[utf8]{inputenc}
\usepackage{kotex}
\usepackage{soul}
\usepackage{comment} 
\usepackage{subfigure} 
\usepackage{lipsum,graphicx,multicol}
\usepackage{enumitem}
\usepackage{balance}
\usepackage{amsmath}
\usepackage[normalem]{ulem}
\usepackage{algorithm,algorithmic}
\usepackage[switch]{lineno}
\usepackage{cite}
\usepackage{stfloats}

\usepackage{booktabs} 
\usepackage{graphicx}
\usepackage{epstopdf}
\usepackage{multicol}
\usepackage{xspace}
\usepackage{comment}
\usepackage{xurl} 

\usepackage{atbegshi}
\usepackage{color}
\usepackage{cite}

\definecolor{mypink1}{rgb}{0.858, 0.188, 0.478}
\definecolor{mypink2}{RGB}{219, 48, 122}
\definecolor{mypink3}{cmyk}{0, 0.7808, 0.4429, 0.1412}
\definecolor{mygray}{gray}{0.6}

\newcommand{\es}[1]{\textcolor{blue}{#1}}

\definecolor{red}{rgb}{.6,0,0}
\definecolor{green}{rgb}{0,.65,0}
\definecolor{brown}{rgb}{0.6,0.6,0}
\definecolor{blue}{rgb}{0,.145,.698}
\definecolor{purple}{rgb}{0,0,0}
\definecolor{cyan}{rgb}{0,.698,.698}
\definecolor{lightgray}{gray}{0.5}
\definecolor{mypink1}{rgb}{0.858, 0.188, 0.478}
\definecolor{mypink2}{RGB}{219, 48, 122}
\definecolor{mypink3}{cmyk}{0, 0.7808, 0.4429, 0.1412}
\definecolor{mygray}{gray}{0.6}

\newcommand\offloadDB{OffloadDB} 
\newcommand\offloadFS{OffloadFS}
\newcommand\offloadPrep{OffloadPrep}

\begin{document}

\title{
OffloadFS: Leveraging Disaggregated Storage for Computation Offloading
}


\author{Sungho Moon,
         Daegyu Han,
         Hera Koo,
         Sangeun Chae, \\
         Duck-Ho Bae, 
         Euiseong Seo~\IEEEmembership{Member, IEEE}, 
         Beomseok Nam~\IEEEmembership{Member, IEEE} 
\IEEEcompsocitemizethanks{\IEEEcompsocthanksitem 
S. Moon, D. Han, H. Koo, S. Chae, E. Seo, and B. Nam  are with Sungkyunkwan University.
D. Bae is with Samsung Electronics. 
}
}

\markboth{ieee transactions on parallel and distributed systems,~Vol.~31, No.~10, november~2019}%
{Shell \MakeLowercase{\textit{et al.}}: Bare Demo of IEEEtran.cls for Computer Society Journals}

\IEEEtitleabstractindextext{%

\begin{abstract}

Disaggregated storage systems improve resource utilization and enable
independent scaling of storage and compute resources by separating storage
resources from computing resources in data centers. NVMe over fabrics (NVMeoF)
is a key technology that underpins the functionality and benefits of
disaggregated storage systems. While NVMeoF inherently possesses substantial
computing and memory capacity, these resources are often underutilized for
tasks beyond simple I/O delegation.  This study proposes {\offloadFS}, a
user-level file system that enables offloaded IO-intensive tasks primarily to a
disaggregated storage node for near-data processing, with the option to offload
to peer compute nodes as well, without the need for distributed lock
management.  {\offloadFS} optimizes cache management by reducing interference
between threads performing distinct I/O operations.  On top of {\offloadFS}, we
develop {\offloadDB}, which enables RocksDB to offload MemTable flush and
compaction operations, and {\offloadPrep}, which offloads image pre-processing
tasks for machine learning to disaggregated storage nodes.  Our evaluation
shows that {\offloadFS} improves the performance of RocksDB and machine
learning pre-processing tasks by up to 3.36$\times$ and 1.85$\times$,
respectively, compared to OCFS2.



\end{abstract}

\begin{IEEEkeywords}
Key-Value Store, LSM Tree, NVMe over Fabric, Disaggregated Storage, \textcolor{purple}{Near-Data Processing}, Compaction Offloading       
\end{IEEEkeywords}
}

\maketitle
\thispagestyle{plain}
\pagestyle{plain}

\IEEEdisplaynontitleabstractindextext
\IEEEpeerreviewmaketitle

\newcommand\blfootnote[1]{%
  \begingroup
  \renewcommand\thefootnote{}\footnote{#1}%
  \addtocounter{footnote}{-1}%
  \endgroup
}

\section{Introduction}\label{intro-section}

\blfootnote{This work has been submitted to the IEEE for possible publication. Copyright may be transferred without notice, after which this version may no longer be accessible.}
NVMe over Fabrics (NVMeoF) has emerged as a leading technology in modern
Storage Area Networks (SANs). NVMeoF enables the transport of NVMe commands
over a network fabric~\cite{nvmeofspec}, allowing remote NVMe SSDs to function
as local block devices. In this regard, NVMeoF shares similarities with
FibreChannel and SCSI-based protocols such as iSCSI and iSER. However, NVMeoF
ensures more efficient access to remote NVMe devices with low latency, as it
bypasses the storage server's OS and I/O stack, supports parallel I/O
operations, eliminates the need for inter-process locking, and minimizes the
number of round trips for a data transfer~\cite{guz2018performance,
xu24lightpool,zieleznicki24}. 




Storage disaggregation enables independent and dynamic scaling of storage
capacity based on workload requirements, preventing overprovisioning and
underutilization of resources.  To enable the pooling of storage devices and
their sharing across multiple compute nodes, NVMeoF-based disaggregated storage
solutions, such as PoseidonOS~\cite{poseidonos}, are being developed for
providing NVMeoF interface out of 
JBOF systems. These NVMeoF JBOF systems create logical volumes of arbitrary
sizes from dozens of physical NVMe SSDs for compute nodes within a
SAN~\cite{legtchenko17}.

NVMeoF is a block-level protocol that allows multiple initiator nodes to access
the same target volume.  When multiple initiators access the volume
simultaneously, IO conflicts may occur. NVMeoF leaves the responsibility for
resolving these conflicts to the file system layer.

NVMeoF JBOF storage nodes are typically equipped with multi-core processors and
a substantial amount of memory capacity.  Although NVMeoF significantly reduces
the demands on computational and memory resources, NVMeoF necessitates DMA
operations, requiring a substantial number of DIMMs on NUMA nodes for high
memory bandwidth~\cite{xeonguideline}.  Additionally, a certain level of
computational power is required to dynamically adjust storage volumes in
response to workload demands and to transmit NVMe commands and data over
standard TCP/IP networks when using NVMe-TCP, one of the transport protocols
within NVMeoF.

However, we note that in a rack-scale SAN, where multiple initiator nodes share
an NVMeoF JBOF node, the resources of the JBOF node are often underutilized.
This indicates that, disaggregated storage introduced to address
the problem of resource underutilization, paradoxically results in the
underutilization of CPU and memory on the storage node.

To mitigate the resource underutilization and enable {\it near data processing}
(NDP), we explore the potential of offloading I/O-intensive tasks to storage
nodes.  Performing I/O tasks on storage nodes not only makes better use of the
storage node's resources but also aligns with the objectives of {\it near-data
processing}.  In-storage processing (ISP), a specific form of NDP that
integrates computational resources into SSDs, brings computation into the
storage where data are stored, reducing the need to transfer large volumes of
data~\cite{choe2017neardata, gao15pact,cao22ishbase}.  In a disaggregated storage
environment with NVMeoF, the CPUs within JBOF nodes can supply the necessary
processing power, which enables ISP without computational SSDs.  

However, offloading I/O-intensive tasks to storage nodes presents two
challenges. First, since a single storage node is shared by multiple compute
nodes, there is a risk that the storage node's resources could become a
bottleneck if multiple compute nodes offload tasks simultaneously. To mitigate
this, it is necessary to selectively offload I/O tasks by monitoring the
resource usage of the storage node.  To the best of our knowledge, there are no
near-data processing studies that employ a mechanism for deciding whether to
offload I/O tasks~\cite{sun2019near,choe2017neardata, gao15pact,cao22ishbase}.
Second, when I/O tasks are offloaded selectively, both the initiator node and
the target node must be able to access the same storage volume simultaneously.
This requirement typically necessitates the use of a shared-disk file system.
However, shared-disk file systems use complex metadata and distributed lock
management protocols to maintain file system consistency. Since I/O offloading
allows only one remote node to perform the I/O tasks permitted by the
initiator, the consistency mechanism can be significantly simplified.


In this study, we develop {\it {\offloadFS}}, a user-level file system designed
for offloading IO intensive tasks to NVMeoF storage nodes. Optionally,
{\offloadFS} allows to offload tasks to peer nodes with the NVMeoF storage
mounted. 
%
%
{\offloadFS} shares some similarities with shared-disk file systems in that
multiple nodes can concurrently perform I/O operations offloaded by the
initiator, but it is specialized for task offloading for applications running
on a single compute node.  That is, the focus of this work is not on
distributed parallel applications where multiple processes access the same
files without coordination.  Therefore, {\offloadFS} is more lightweight and
offers simpler consistency guarantees across NVMeoF initiator and target nodes
compared to conventional shared-disk file systems.

To showcase the applicability of {\offloadFS}, we develop two applications on
top of it. First, we develop {\offloadDB}, a variant of RocksDB that offloads
background compaction tasks. Second, {\offloadPrep} is an image processing
library for machine learning (ML) pipelines that enables offloading of image
pre-processing tasks to disaggregated nodes.


This study makes the following contributions.

\begin{itemize}
[leftmargin=*]
\setlength\itemsep{-0.2em}

\item Given that the connectivity of NVMeoF leaves the compute and memory
resources of NVMe-oF target nodes underutilized, this study explores the
potential for opportunistically harnessing these unused resources, and develops
a file system for offloading IO intensive tasks to a storage node for near-data
processing. 


\item We design and implement {\offloadDB} atop {\offloadFS}.  {\offloadDB}
enables the initiator to offload background compaction tasks to the target
node, such that they perform different types of IO operations. This design
isolates background IOs from foreground IOs, allowing each node to cache
specific data required for its own IO tasks. 




\item We develop a preprocessing library, {\offloadPrep}, on top of
{\offloadFS}, to offload ML pre-processing tasks to remote nodes. Using
{\offloadPrep}, ML preprocessing tasks can benefit from
lightweight concurrency control, even when offloading to multiple peer compute
and storage nodes.


\end{itemize}

The rest of this paper is organized as follows. In Section
\ref{background-section}, we present the background and motivation of this
study.  In Sections~\ref{offloadFS-section}, \ref{offloadDB-section} and \ref{other-section}, we
present the design of {\offloadFS} and its applications.  In
Section \ref{experiments-section}, we present the performance evaluation.
Finally, we conclude in Section~\ref{conclusion-section}.

\section{Background}\label{background-section}

\subsection{NVMe over Fabrics}


NVMeoF (NVMe over Fabrics) is a network protocol specification designed to
enable the dispatch of NVMe commands over a network fabric~\cite{nvmeofspec}.
Leveraging RDMA, NVMeoF provides direct access to remote NVMe storage, as if it
were a local device.  NVMe specification~\cite{nvmeofspec} defines an interface
for host software to communicate with non-volatile memory subsystems over a
variety of memory-based transports and message-based transports, including RDMA
and TCP transports.  

NVMeoF allows the NVMe host driver on the initiator node to pass NVMe commands
to either RDMA or TCP transport rather than the PCIe transport. With the RDMA
transport, NVMe commands are transferred to the target node using messages, and
the data are transferred using memory semantics via RDMA.  Upon receiving the
NVMe command, the RNIC at the target node signals the target's NVMe RDMA
transport. Then, a {\it reactor} running on the target node processes the NVMe
command, constructs a response, and passes it back to the RDMA transport. This
design allows NVMeoF to bypass the conventional system software stack,
resulting in increased bandwidth, reduced latency, and lower CPU utilization
on the target node.

\subsection{Access to NVMeoF SSDs via File Systems}\label{background:filesys-section}

\begin{figure}[!t]
\centering
    \includegraphics[width=1\columnwidth]{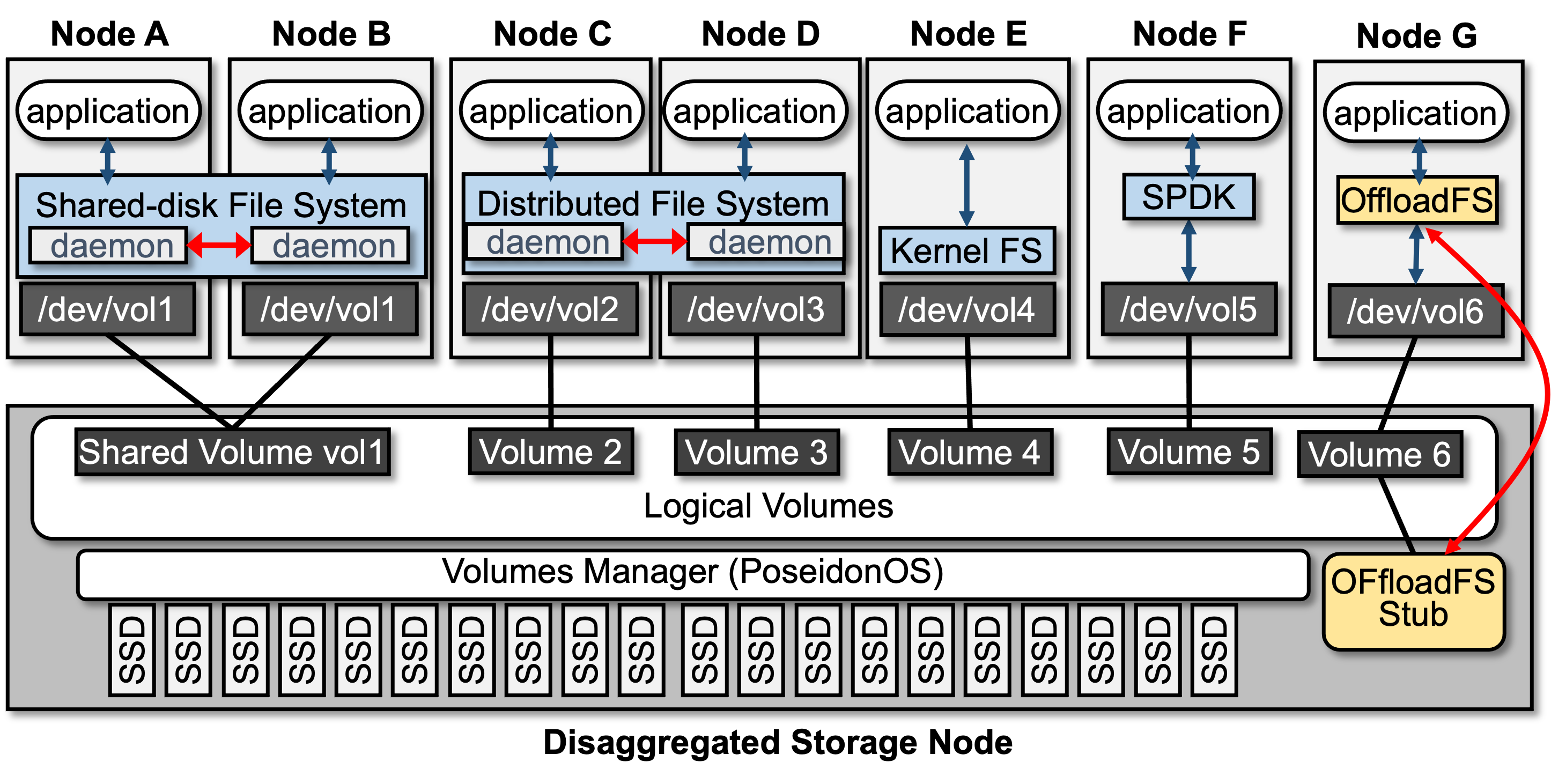}
    \caption{Various File System Configurations with JBOF}
    \label{fig:JBOF}
\end{figure}

\begin{figure}[!t]
    \centering 
        \includegraphics[width=0.85\columnwidth]{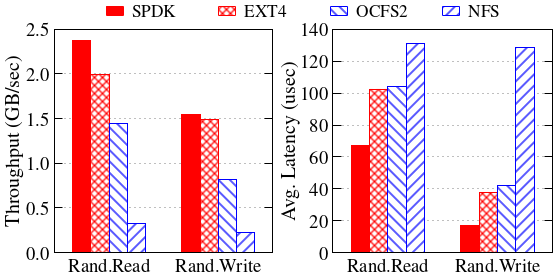}
    \caption{Performance with Various File Systems and SPDK}
    \label{fig:fio}
\end{figure}

With the emergence of NVMeoF, software-defined storage solutions, such as
PoseidonOS~\cite{poseidonos}, are being developed to provide storage pools
using the NVMeoF protocol. PoseidonOS creates and manages logical volumes from
a pool of NVMe SSDs. These logical volumes are exposed as block devices to
initiators, in compliance with the NVMeoF specification. As a block
device-level protocol, NVMeoF leaves the file system management to the client.  

Figure~\ref{fig:JBOF} shows various file system configurations using NVMeoF.
In shared-disk file systems, multiple initiators share the same physical
NVMe volume by mounting it simultaneously on multiple servers. To manage
concurrent accesses, each initiator node requires a daemon process, and these
processes communicate with each other via RPC to ensure consistent file access.
Shared-disk file systems are used when multiple nodes need to share the same
data, making them commonly used in parallel systems such as high-performance
scientific applications, Oracle RAC, and home directories for multi-user
cluster systems.

In distributed file systems, each node has its own storage volume,
accessible only by itself. The file system then abstracts and manages the
distributed storage devices across multiple nodes for
large scale divide-and-conquer processing. 


In addition to such scale-out applications, there exist numerous applications
that do not require a clustered file system.  For example, RocksDB and
single-node DBMS such as MySQL do not share files with peer compute nodes. For
these applications, storage can be accessed through a non-cluster file system.
This study focuses on those single-node applications and enables them to offload
I/O tasks primarily to target nodes, with the option to include peer compute
nodes as well, for applications running on a single initiator node that
accesses disaggregated storage.



%

\subsection{Offloading with Shared-Disk File Systems}\label{background:shared_filesys_perf-section}

Offloading has been extensively studied in cloud computing
environments~\cite{wang19edge}. Most of the studies focus on scheduling
techniques for selecting which remote nodes or accelerators to use.  In
contrast, near-data processing performs computations at the location of the
data, rather than dynamically deciding which node should handle each
computation. Near-data processing techniques commonly pre-determine which
specific computations will be performed where the data is located.  If the
accelerator, such as a computational SSD, is busy, offloaded tasks simply
wait~\cite{kim2021linefs,cao22ishbase,zhang2020fpga}. 


For selective offloading, i.e., to be able to choose whether or not to offload
an I/O task to a remote helper node, the compute node requesting the offloading
and the helper node must be able to access the same file. Therefore, a large
number of previous I/O offloading studies use shared-disk file systems to
offload computation to FPGA~\cite{sun2019near,sun2020fpga,zhang2020fpga},
smart NIC~\cite{kim2021linefs}, or even computational SSD on a single
host~\cite{lim21icce}, despite computational SSDs enable applications to
offload computations by passing block addresses to the device for access. 


For selective offloading, i.e., to be able to choose whether or not to
allow multiple servers to concurrently access the same storage, shared-disk
file systems require complex metadata management and a distributed lock manager
to ensure data consistency and cache coherence. Unfortunately, the overhead
associated with these mechanisms is significant, which results in lower
performance compared to non-cluster file systems.

To illustrate this overhead, we mount a disaggregated NVMe-oF volume on a single
initiator (compute) node using both  cluster file systems and local file
systems, and run Random Read and Random Write workloads using the FIO benchmark.
As shown in Figure~\ref{fig:fio}, EXT4 achieves substantially higher throughput 
and slightly lower latency than OCFS2. Notably, this experiment involves only a 
single client node and therefore does not exhibit conflicting accesses. 
Nevertheless, OCFS2 still incurs the overhead from distributed locking
and metadata coordination required by the DLM~\cite{wang23cfs,park26lockify}.

\subsection{Resource Under-Utilization on Target}\label{underutil-section}

\begin{figure}[!t]
\subfigure[Throughput Varying Number of Reactors]{
    \includegraphics[width=0.55\columnwidth]{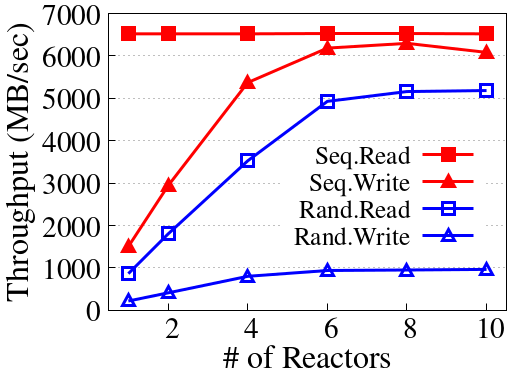}
	\label{fig:varying_reactor}
}
\subfigure[Memory Usage Varying Number of NVMe Devices]{
    \includegraphics[width=0.38\columnwidth]{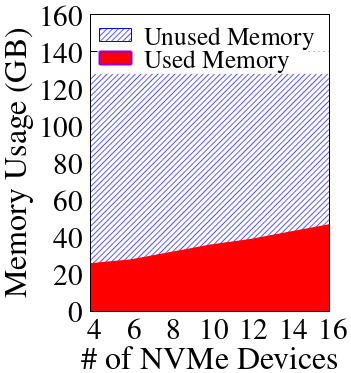}
	\label{fig:varying_nvme}
}
\caption{Resource Under-utilization on Storage Node}
\label{fig:resource_util}
\end{figure}

Unlike conventional network storage protocols, NVMeoF reduces the need for
computational resources on the storage node~\cite{guz2018performance}. In the
experiments shown in Figure~\ref{fig:resource_util}, we run FIO benchmarks
while varying the number of reactors and NVMe SSDs on {\it
PoseidonOS}~\cite{poseidonos}, an NVMeoF storage management system developed
jointly by Inspur and Samsung. 
From this experiments, we made three observations.

First, CPU utilization of the target node is very low.  Even with 6 reactors
(cores), the throughput of an array of 16 NVMe SSDs saturates.  This is a well
known fact because the RDMA transport allows bypassing the conventional
software stack.  Unlike native NVMeoF devices, PoseidonOS uses the CPU cycles
to monitor WQEs (Work Queue Entries) posted by the initiator and perform
address translation from logical volumes to physical block addresses.
Nonetheless, it is possible to achieve the saturation point for an array of
NVMe SSDs throughput with a limited number of reactors~\cite{spdkreport23}.  In
a native NVMeoF environment, the throughput of an array of NVMe SSDs can be
saturated with fewer reactors.

Second, the memory utilization on the target node is low. NVMeoF utilizes
memory for DMA, but the required size of the DMA buffer is not large.  In the
experiments shown in Figure~\ref{fig:varying_nvme},  we measure the memory
usage with varying the number of NVMe SSDs.  When the number of NVMe devices
increases from 4 to 16, the memory usage increases by 21~GB. This indicates
that the DMA buffer size required for each NVMe SSD is only about 1.75~GB,
i.e., the required memory size of storage nodes is not substantial. However, it
should be noted that it is a common practice to install more memory than
required in order to increase DMA performance by utilizing a larger number of
DIMMs~\cite{xeonguideline}.  Using a single DIMM on each memory controller, for
example, results in only 35\% of the maximum system memory
bandwidth~\cite{xeonguideline, mempoprule}.

Third, unlike Direct-Attached Storage (DAS), NVMeoF introduces network
communication overhead, posing issues not only with network
latency~\cite{guz18nvmeof} but also consuming significant network bandwidth.
That is, the IO bandwidth is limited not by the bandwidth of the storage array
but rather by the network bandwidth.  It is noteworthy that storage bandwidth
can be expanded relatively easily by adding more
disks~\cite{thereska2013ioflow}.  However, expanding network bandwidth requires
adding or upgrading network interfaces, upgrading network switches, and
installing more network cables.  A cost-effective solution to this challenge is
near-data processing, which not only saves network bandwidth but also reduces
latency and the load on network resources.

To address these challenges, this study investigates the development of an
SPDK-based user-level file system optimized for offloading I/O-intensive tasks
to the NVMeoF target node, such that the idle resources of NVMeoF target nodes
are opportunistically utilized.

\subsection{Target Application: LSM Trees}

Log-structured merge trees (LSM tree)~\cite{O'Neil1996LSM}, designed with a
focus on fast write operations, are widely adopted as the core storage engine
for popular key-value stores such as RocksDB~\cite{RocksDB},
LevelDB~\cite{LevelDB}, Cassandra~\cite{Lakshman2010cassandra}, and
HBase~\cite{HBase}.  Key-value stores constructed atop LSM trees employ
multiple worker threads to manage multi-level indexing structures.
Specifically, foreground threads handle client requests while background
threads reorganize the multi-level index by merge-sorting SSTables to improve
search performance.

The background threads, called {\it compaction threads}, often interfere with
foreground threads that serve  client requests. In order to avoid the
interference of background worker threads with foreground threads, most LSM
tree implementations often pause foreground threads until background threads
finish the compaction. The issue known as the {\it write stall} problem is
rooted in the hierarchical structure of LSM trees, periodic compaction, and the
performance gap between memory and storage
devices~\cite{LevelDB,Balmau19silk,kim22listdb}.

\subsubsection{Compaction Offloading}\label{background:offloading-section}

To mitigate the write stall problem, numerous studies offered different
perspectives~\cite{ahmad2015compaction,Balmau19silk,Balmau2017Flodb,
bindschaedler2020hailstorm,Dong21rocksdb,Kaiyrakhmet19SLMDB,Kim20Bolt,
Lepers19kvell,Raju2017Pebblesdb,sun2019near,sun2020fpga,yao20matrixkv,
zhang2020fpga}.  
%
%
Some of these prior studies identified CPU resource shortages as the cause of
write stall and explored the potential of offloading compaction tasks to
accelerators, such as FPGA~\cite{lim21icce,sun2019near,sun2020fpga,
zhang2020fpga}, GPU~\cite{sun23GLSM}, and DPU~\cite{ding23DComp}.  There are
also some proposals that utilize the computing resources of idle remote 
nodes~\cite{ahmad2015compaction,cao22ishbase,bindschaedler2020hailstorm,
li2021elastic}. 

One of the challenges in compaction offloading is the sharing of SSTables
between the host and a remote node.  Most existing studies have addressed this
challenge by using distributed file systems. For example, Meta implemented a
disaggregated compaction method for RocksDB using {\it Tectonic} distributed
file system~\cite{dong23}.  Ahmad et al.~\cite{ahmad2015compaction} proposed
offloading compaction tasks for HBase to remote idle nodes using HDFS.
Similarly, Bindschaedler et al.~\cite{bindschaedler2020hailstorm} introduced
Hailstorm, a distributed file system designed to enable compaction offloading
for MongoDB.  Li et al.~\cite{li2021elastic}, proposed {\it FaaS compaction},
where compaction tasks in TerarkDB~\cite{terarkdb} are offloaded to a remote
FaaS cluster.  The FaaS cluster copies victim SSTables from the shared data
store to a local EXT4 file system and then returns the output SSTables
back to the shared data store. By distributing compaction tasks across multiple
nodes, they achieve load balancing and improve throughput under skewed workloads.
There are also studies that use shared-disk file systems as an
alternative to distributed storage systems. Lim et al.~\cite{lim21icce} used
OCFS2 to enable in-storage compaction with OpenSSD. Ding et al.~\cite{ding24}
developed DALFS, a DPU-aware and LSM-specialized file system that offloads
compaction tasks to the DPU. Zhang et al.~\cite{zhang2020fpga} proposed 
accelerating compaction tasks using FPGA, where their storage system, called
X-engine\cite{huang19}, is tightly coupled with the FPGA.

\subsubsection{Cache Pollution}\label{pollution-section}



Cache plays a crucial role in LSM trees, and there have been various studies
examining cache efficiency in LSM trees~\cite{teng18tos,wu20ackey,yang20vldb,
wang24icde}.  However, to the best of our knowledge, no prior research has
investigated the cache pollution problem caused by compaction. 

I/O access in the LSM-tree exhibits a distinct pattern.  Background compaction
threads sequentially read victim SSTables and then write new SSTables in the
same sequential fashion. After victim SSTables are read, they are soon deleted
after compaction. This makes caching victim SSTables ineffective and wasting
valuable cache space. 

Not only are the blocks of soon-to-be-deleted SSTables stored in the cache, but
also the blocks of the newly generated SSTables contribute to the cache
pollution problem because they are proactively generated by background
compaction threads. Compactions are initiated not because clients access new
SSTables, but because background threads anticipate that search performance
will benefit from compaction. Even though RocksDB provides an option not to
cache newly created SSTables in its user-level block cache, file blocks are
still cached in the file system, resulting in inefficient memory usage.
Offloading compaction tasks effectively addresses such cache pollution issues.

\subsection{Target Application: ML  Pre-Processing}

In DNN training pipelines, input files are retrieved from storage and then {\it
pre-processed} in memory. For instance, in computer vision, image files are
often rotated, cropped, flipped, and resized. These pre-processing tasks are
typically performed on the compute node in a pipelined fashion while the GPU
trains the model using the pre-processed input data.  If the pre-processing
throughput does not match the GPU training throughput, the {\it data stall}
problem occurs and results in reduced GPU utilization. For instance, Mohan et
al.~\cite{mohan21vldb} showed that the pre-processing accounts for up to 65\%
of the training time.  

To address the data stall problem, various offloading
techniques~\cite{audibert23socc,graur22cachew,graur24pecan,mohan21vldb,wang24,
kim24fusionflow} have been explored to improve the pre-processing performance.
In particular, {\it tf.data.service}~\cite{audibert23socc} built on top of
{\it tf.data} in TensorFlow executes input data pre-processing pipelines in a
distributed manner.  Cachew~\cite{graur22cachew} is another service built on
tf.data in TensorFlow, which conrols the number of distributed input data
workers in serverless computing environment.  Sophon~\cite{wang24} proposes to
selectively offload only a portion of the pre-processing. These works
propose to offload pre-processing tasks to remote CPU servers. In contrast,
Pecan~\cite{graur24pecan} and FusionFlow~\cite{kim24fusionflow} propose to
schedule data workers across both DNN training nodes with GPUs and remote CPU
nodes. 
Unlike these previous studies, which use shared-disk file systems or
distributed object stores and focus on leveraging idle resources of remote
compute nodes, our study complements their work by showing the benefits of
performing pre-processing on storage nodes, adding a new dimension to the
existing research.

\section{Design of \offloadFS}\label{offloadFS-section}

\begin{figure}[!t]
    \centering 
    \includegraphics[width=1.0\linewidth]{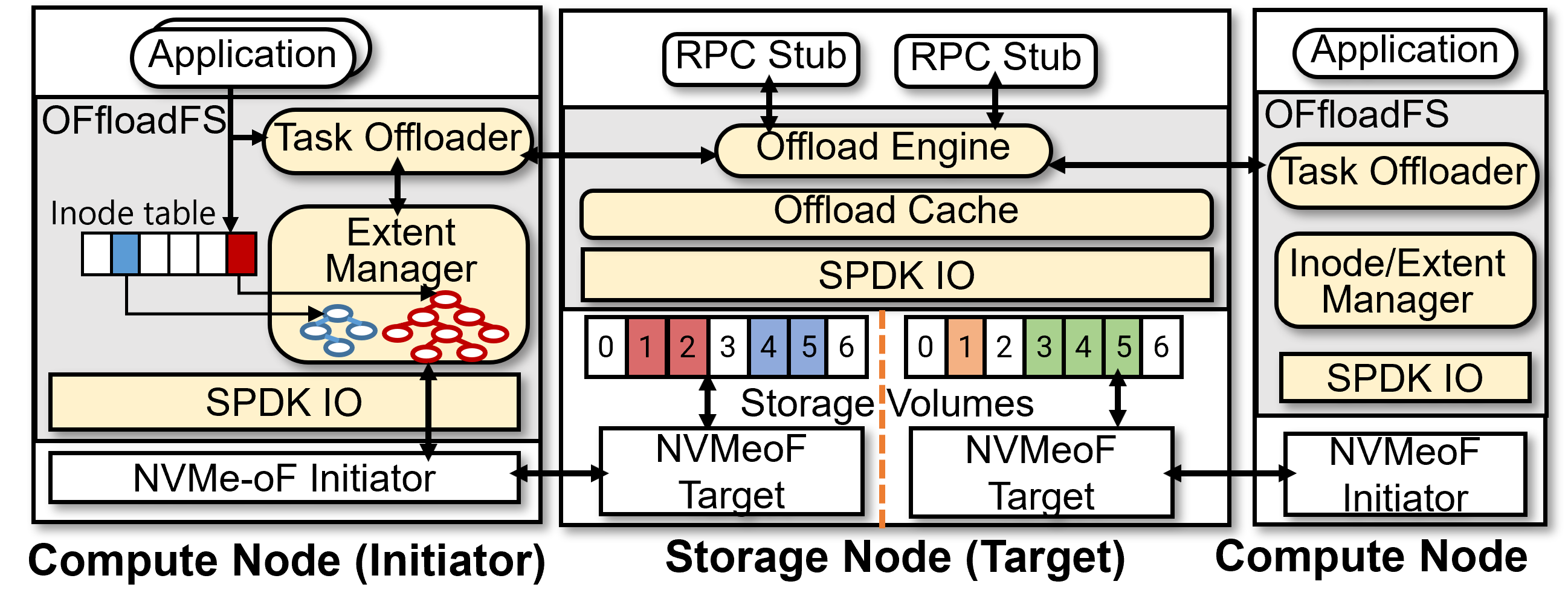}
    \caption{ {\offloadFS} Architecture}
    \label{fig:arch}
\end{figure}


{\it {\offloadFS}} is an initiator-centric, user-level file system that
offloads I/O-intensive tasks to a remote node by enabling user-defined
functions to be executed via gRPC.  The remote node is granted access and
performs I/O operations only on the blocks authorized by the initiator.  In
this section, we first describe offloading to target storage nodes, primarily
to utilize the idle resources of storage nodes. Then, in
Section~\ref{peer_offload-section}, we discuss how peer initiator nodes perform
offloaded tasks by leveraging the distance connectivity of NVMeoF.


Figure~\ref{fig:arch} shows the architecture of {\offloadFS}.  On the initiator
node, {\offloadFS} manages the {\it inode table} and extents.  The {\it Extent
Manager} allocates extents, manages free space and per-file extent trees, and
performs defragmentation. 

One of the key components of {\offloadFS} is {\it Task Offloader}, which
offloads IO tasks to the target node via gRPC.  When it's beneficial to execute
IO on the target node according to the application's judgment, distributed RPC
stub and skeleton codes need to be implemented in the application level. The
RPC stub on the target node perform IOs using \texttt{offload\_read()} and
\texttt{offload\_write()} provided by {\offloadFS}.  These APIs use SPDK to
read and write data to the block addresses received as arguments.  By using
file system APIs, {\offloadFS} ensures that only authorized blocks can be
accessed, unlike computational SSDs. In addition, by caching the read blocks
in the {\it Offload Cache}, it reduces the number of I/O operations if the
same blocks are accessed repeatedly. 


The offloaded IO tasks, implemented as gRPC stub codes, can (i) read and
compute on existing extents, (ii) update existing extents, and (iii) return
task results. These operations can be executed individually or in any
combination, depending on the gRPC stub implementation.  Offloading of
directory-related IO tasks is not supported. Additionally, IO tasks that
require access or modifications of the inode table and extent trees such as
{\it truncate}, {\it fallocate}, {\it stat}, and {\it ioctl} cannot be called
by the offloaded task on the target node. This restriction exists because the
inode table and extent trees are solely managed by the initiator node. 
For instance, if the initiator node intends to offload a task that creates a
new file, a new inode must be allocated on the initiator node before
offloading. Similarly, to offload a task that writes a new extent, the
initiator node must preallocate the necessary blocks. Only after the inode
table and extent trees are prepared can the physical block addresses be
provided to the target node as arguments in the RPC function.

Due to this characteristics, which we refer to as the {\it initiator-centric
block management policy}, {\offloadFS} prevents metadata consistency issues
despite two nodes access the same storage volume.  Specifically, the initiator
node keeps track of the the addresses of blocks that offloaded IO tasks are
accessing. The initiator is not allowed to access those blocks until receiving
a notification that the offloaded task has been completed. That is, there will
be no concurrent access to blocks that cause conflict. Hence, the target node can
read and write without the need to obtain distributed locks.



\subsection{Initiator-centric Cache Coherence}\label{caching-section}

{\offloadFS} does not provide caching on the initiator node. To improve I/O
performance, applications may implement application-level block caching as
needed.  On the target node, the {\it Offload Engine} provides caching to make
use of the underutilized memory. Specifically, if offloaded tasks call
\texttt{offload\_read()}, the Offload Engine  searches the {\it Offload Cache},
and if the requested block is not found, it reads the block from NVMe SSDs,
copies it to the cache, and then pins it until the offloaded task finishes.

The initiator node can write directly to NVMe SSDs without going through the
Offload Engine.  As a result, blocks cached in the Offload Cache can become
stale, leading to potential cache coherence issues. The traditional method to
address this problem is to invalidate the corresponding block in the cache when
a write occurs. However, as previously discussed, traditional shared-disk file
systems suffer from cache coherence mechanisms as they exchange a large number
of messages between nodes.  This also goes against the intent of NVMeoF, which
aims to minimize the use of computation resources on the target node.

{\offloadFS} assigns the responsibility of maintaining cache coherence to the
application. In other words, just as the initiator specifies the block address
for I/O when offloading a task, it also decides whether to read the block from
the Offload Cache or read directly from NVMe, bypassing the cache. Thereby, it
eliminates the need for messsage exchanges to maintain cache coherence.

Alternatively, the initiator node checks the file's modified time and passes it
as an argument to the Offload Engine. If the Offload Engine detects that a
cached block is older than the file's modified time, it bypasses the cache and
reads the block directly from NVMe. This coarse-grained cache coherence
protocol may unnecessarily read blocks from NVMe even if they have not been
modified. But, {\offloadFS} is not designed as a general-purpose shared-disk
file system for use cases where multiple processes simultaneously access files.
Instead, it is intended for applications where  their initiator nodes are
entirely responsible for offloading I/O.  Therefore, using metadata to manage
whether blocks have been modified at the block level might not provide
significant performance benefits relative to its complexity.

\subsection{Multi-Tenancy Support}


In a disaggregated storage architecture, a single storage node is shared by
multiple compute nodes, allowing for higher resource utilization and
scalability.  Therefore, offloading IOs to the storage node from multiple
nodes can lead to resource contention. To prevent this, {\offloadFS} provides
two options. 

The first option is the reactive {\it threshold-based offloading control}.  If
the storage node's resource usage (e.g., CPU utilization) surpasses the
predefined threshold, the storage node rejects new incoming offloading
requests. This is a protective method to prevent the storage node from becoming
overloaded, which could degrade performance.  If an offloading request is
rejected, the Task Offloader on the initiator node will execute the task
immediately on the initiator node itself, rather than waiting for the target
node to become available.

In the second option, a proactive one, the Offload Engine circulates a fixed
number of tokens to the Task Offloaders on initiator nodes, and a Task
Offloader holding a token can send an offload request within a given time
frame. These tokens have a time-to-expire, and if a compute node does not
perform offloading within a certain time, the token is reclaimed, enabling
another compute node to execute offloading.  The token-based policy, unlike
methods that monitor the storage node’s CPU load, has the disadvantage of
requiring tuning of the token’s time-to-expire. However, it has the advantage
of ensuring fairness among Task Offloaders.

\subsection{Offloading to Peer Initiator Nodes}\label{peer_offload-section}

NVMeoF storage can be mounted on multiple initiator nodes. When multiple compute nodes
mount the same NVMeoF target volume, they can access the same volume via block addresses. In this case,
{\offloadFS} can offload tasks to a peer compute node in addition to the target storage node
as long as the block addresses are known. 
This makes {\offloadFS} similar to conventional shared-disk
file systems, but it differs in that consistency is not managed through
communication between daemon processes. Instead, the initiator grants block
access permissions via RPC.  When a volume is shared among multiple initiator
nodes, the application, rather than {\offloadFS}, is responsible for deciding
which peer initiator node should run the offloaded tasks.

\section{Design of {\offloadDB}}\label{offloadDB-section}

\begin{figure}[!t]
    \centering 
    \includegraphics[width=0.9\linewidth]{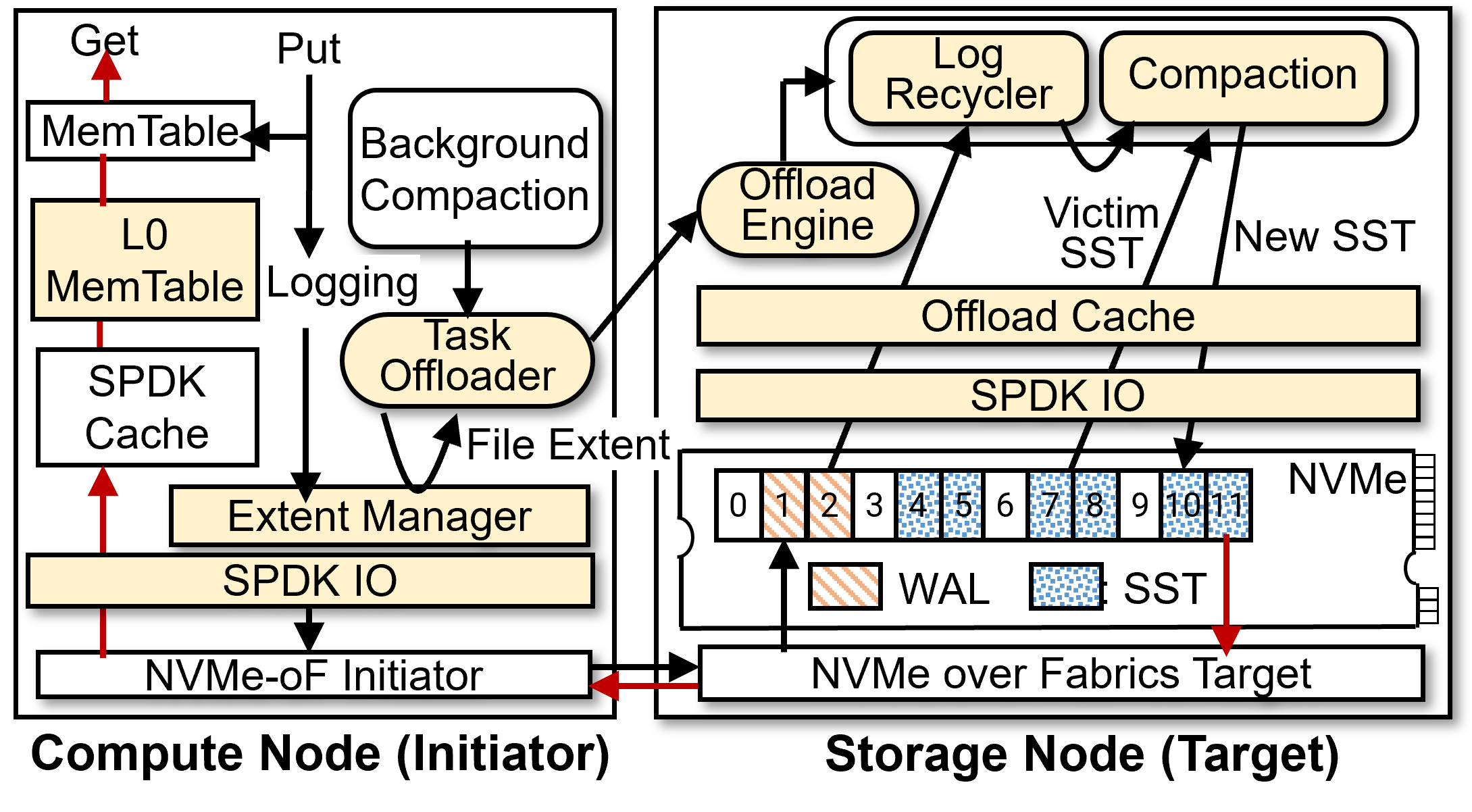}
    \caption{ {\offloadDB} Architecture}
    \label{fig:offloaddb}
\end{figure}

{\offloadDB} is a variant of RocksDB that enables compaction to be performed on
the target node using {\offloadFS}.  By offloading compaction tasks to the
target node, {\offloadDB} reduces network traffic, utilizes high bandwidth of
an array of NVMe's, and avoids cache pollution.  

\subsection{Compaction Offloading}

In RocksDB, there are four types of IO operations: (1) logging for individual
write requests, (2) creating L0 SSTables by flushing MemTables, (3)
merge-sorting SSTables during compaction, and (4) updating the MANIFEST file to
commit SSTables created and deleted during compaction.  In {\offloadDB}, two
background IO operations, i.e., flush and compaction, are offloaded to the
target node, but the other two IO operations, i.e., reading and writing of WAL
and MANIFEST files, are handled by the initiator, as they are frequently
accessed by foreground threads.  As shown in Figure~\ref{fig:offloaddb}, when a
client inserts a key-value into a MemTable, {\offloadDB} requests the Extent
Manager to allocate a new block (Block 2 in the example) for WAL if there is no
space available in pre-allocated blocks. It then writes a log entry using SPDK.

The MANIFEST file is responsible for managing the hierarchical structure of
valid SSTables. Using the MANIFEST file, the background compaction thread
examines the number of SSTables in each level and selects victim SSTables for
the compaction. Then, the compaction thread submits a compaction offloading
request to the Task Offloader, which requests the Extent Manager to create and
allocate storage space for the output SSTables (in the example, blocks 8 and 9
are allocated for output SSTables).  Given that the size of new output SSTables
cannot exceed the size of victim SSTables, the Extent Manager allocates storage
space equivalent to the size of the input SSTables.  The block address
information and metadata about the selected victim SSTables are sent to the
Offload Engine on the target node via RPC.

Upon receiving the block addresses of the victim and output SSTables, the
offloaded compaction task, i.e., RPC stub, reads the victim SSTables (stored in
blocks 4 and 5) and creates new output SSTabls (in blocks 8 and 9). Once the
new SSTables are written, the Offload Engine sends back the metadata of the
new SSTables to the initiator.  The output SSTables may end up smaller in size
than the input SSTables. As a result, the Offload Engine returns information
about the unused blocks to the Block Allocator, to utilize them for future
allocations.

Finally, the Task Offloader on the initiator node updates the MANIFEST
file by replacing the victim SSTables' information with that of the new
SSTables.  If a system crashes after allocating blocks for new SSTables
but before updating the MANIFEST file, the data written to those
allocated blocks is considered garbage and the allocated blocks are reclaimed.
Therefore, updating the MANIFEST file serves as the commit mark for the
compaction operation, as in vanilla RocksDB.


\begin{figure}[!t]
    \centering 
    \includegraphics[width=0.9\linewidth]{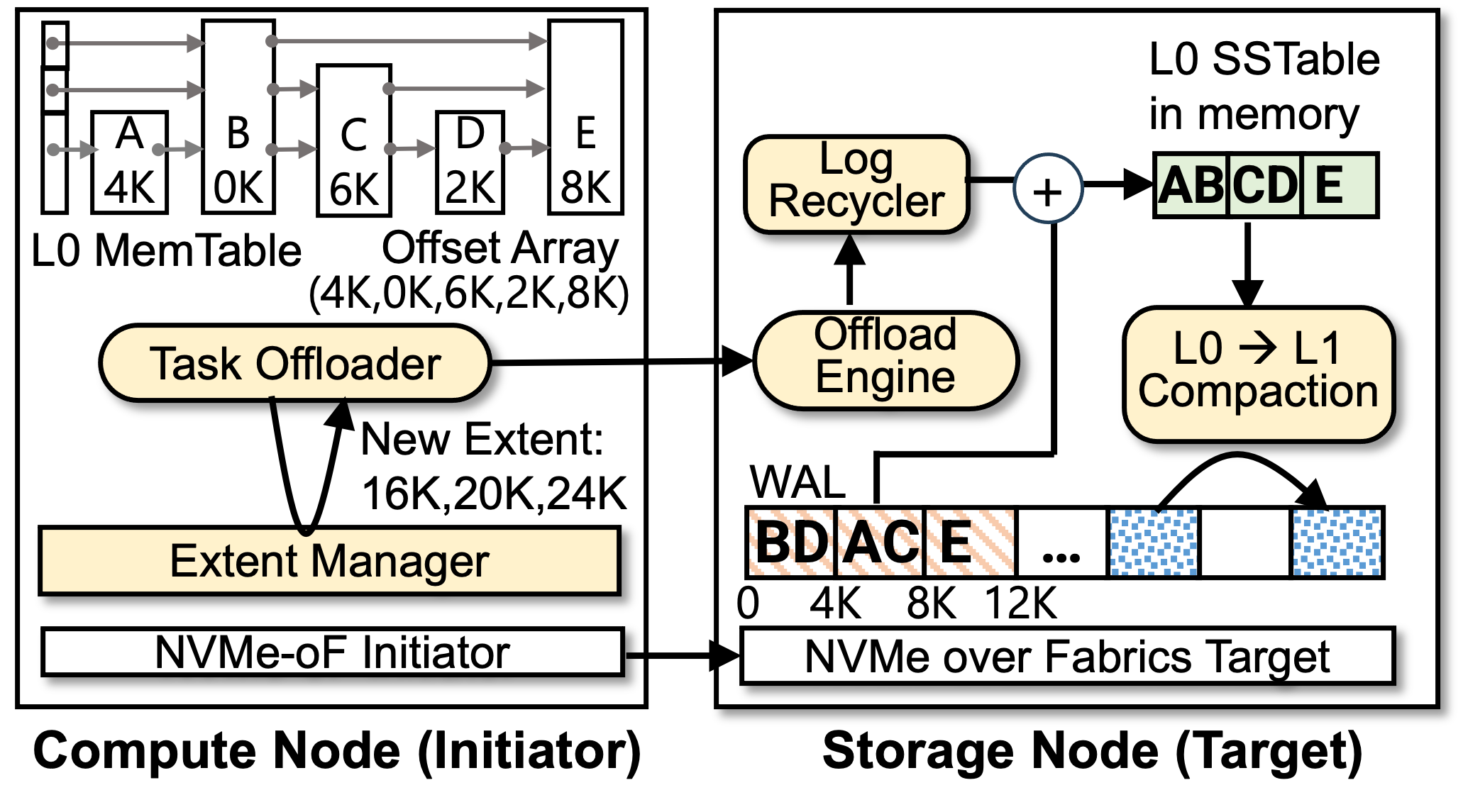}
    \caption{Log Recycling}
    \label{fig:logrecycle}
\end{figure}

\subsection{Log Recycling}\label{log_recycle-section}

In vanilla RocksDB, when a client insert a key-value pair, it is indexed in the
MemTable and also appended to the WAL file.  When NVMeoF storage is used, the
NVMe device that stores the WAL resides on the target node. Therefore, the
key-value pair is transmitted to the storage node through the network.  Later,
when a MemTable is flushed, its key-value pairs are sent to the target node
again, in the form of an L0 SSTable.


Sending the same key-value pairs from the initiator node to the target node
twice (first for writing to the WAL and then for creating an L0 SSTable) can
result in the wastage of network bandwidth. To address this problem,
{\offloadDB} employs {\it Log Recycling} to reduce the duplication of data
transfers and optimize network bandwidth usage. The key idea of Log Recycling
is similar to those of TRIAD~\cite{Balmau2017TRIAD} and
ListDB~\cite{kim22listdb}.  With Log Recycling, the initiator does not transmit
immutable MemTables to the target node.  Instead, {\offloadDB} offloads the
task of performing Log Recycling to the target node. For the Log Recycler RPC
stub, {\offloadDB} sends the block addresses of the WAL to the Log Recycler RPC
stub. Then, the Log Recycler on the target node reads the WAL to reconstruct L0
SSTable.

The WAL contains the key-value pairs that would have been stored in the L0
SSTable, but the key-value pairs are not sorted. To address this problem,
{\offloadDB} stores the file offsets of log entries along with the key-value
pairs in the MemTable.  When a MemTable becomes immutable, i.e., the immutable
MemTable is ready to be flushed to the storage as an SSTable, a background
compaction thread on the initiator node collects these offsets by traversing
the bottom level SkipList pointers of the MemTable.  The offset array is then
sent to the Log Recycler on the target node via RPC.  The offset array is
small, containing only the file offsets without the actual key-value records.
Using the offsets and the WAL blocks, the Log Recycler can read key-value pairs
in a sorted order and reconstruct the sorted array of key-value pairs.
Compared to the conventional L0 SSTable flush, Log Recycling incurs negligible
communication overhead.


Figure~\ref{fig:logrecycle} shows how Log Recycling flushes an immutable
MemTable. When the MemTable is to be flushed, the offset array stored in the
MemTable is passed to the Log Recycler RPC call as an argument.  Additionally,
storage space for the output SSTable is allocated from the Extent Manager, and
the block addresses (16K, 20K, 24K in the example) are provided as arguments to
the RPC call.  Then, the offloaded Log Recycler reads the WAL sequentially
(i.e., B $\rightarrow$ D $\rightarrow$ A $\rightarrow$ C $\rightarrow$ E), and
rearranges those key-value pairs according to the order specified in the offset
array (i.e., WAL[4K]=A $\rightarrow$ WAL[0]=B $\rightarrow$ WAL[6K]=C
$\rightarrow$ WAL[2K]=D $\rightarrow$ WAL[8K]=E). 


\subsubsection{L0 Cachce}\label{l0caching-section}

L0 SSTables, unlike those in upper levels, have overlapping key ranges, which
is known to hurt search performance. Therefore, when a small number of L0
SSTables accumulate, background compaction threads merges them into L1 SSTables
with the highest priority.  During its short lifespan, L0 SSTables are needed
for two purposes.  First, it is read by foreground queries on the initiator.
Second, it is read by background compaction threads on the target.  
For foreground queries, it is more efficient to keep the immutable MemTables in
the initiator's cache, which we refer to as {\it L0 cache}. These MemTables
remain in the cache until they are deleted by L0$\rightarrow$L1 compaction.
Since their lifespans are short, the number L0 SSTables is limited.  L0
SSTables, i.e., immutable MemTables, are evicted from the L0 cache upon
deletion in the MANIFEST. That is, when an L0 SSTable is merged into L1
SSTables, it is no longer required by foreground queries, and removed from the
cache.

When the MemTable size is 64MB, an L0 cache of 3GB can hold as many as 48
immutable MemTables. 48 is a sufficiently large number because L0 $\rightarrow$
L1 compaction is initiated when the number of L0 SSTables exceeds 10 in
RocksDB. Therefore, unless there is an extreme shortage of memory, and provided
that compaction is sufficiently fast, {\offloadDB} does not block incoming
writes, i.e., write stall does not occur.

With L0 cache,  Log Recycling can defer the reconstruction of L0 SSTables until
the L0$\rightarrow$L1 compaction is triggered.  This optimization eliminates
the necessity of storing L0 SSTables on storage because L0 SSTables is needed
only for L0$\rightarrow$L1 compaction.  If the L0 cache is not used, the L0
SSTable reconstructed by the Log Recycler must be stored on storage, and
foreground queries need to read the L0 SSTable from disaggregated storage,
which is highly inefficient.

\section{{\offloadPrep}}\label{other-section}

While we made significant modifications to RocksDB’s compaction and MemTable
flush operations to fully leverage the features of {\offloadFS}, such extensive
modifications are not always necessary to use {\offloadFS}.  To demonstrate
that other applications can improve performance without modifying data
processing logics, we implement {\offloadPrep}, a library that offloads image
pre-processing operations atop {\offloadFS}.

As discussed in Section~\ref{background-section}, computer vision applications
perform pre-processing operations, such as resizing, cropping, flipping, and
rotation. Previous studies~\cite{audibert23socc,graur22cachew,graur24pecan,
mohan21vldb,wang24,kim24fusionflow} showed that offloading image pre-processing
tasks can utilize idle CPUs or GPUs and reduce DNN training time, which
mitigates the data stall problem.  Our focus, however, is not to demonstrate
that pre-processing can reduce the overall training time, but rather to show
that the pre-processing time can be reduced if the pre-processing tasks are
performed through {\offloadFS} on a remote node. 

{\offloadFS} does not compete with previous pre-processing
schemes~\cite{graur24pecan,kim24fusionflow}, but it provides an opportunity to
access training datasets from any node without complex file system consistency
mechanisms. In this regard, our approach is complementary to existing
methods~\cite{audibert23socc,graur22cachew,mohan21vldb}. Specifically,
{\offloadPrep} does not provide scheduling features proposed in
previous works.  Instead, it relies on {\offloadFS}'s load control mechanisms,
such as rejecting offload requests when CPU usage is high or managing
offloading using tokens.


\section{Performance Evaluation}\label{experiments-section}

\subsection{Experimental Setup}\label{setup-section}

We run experiments on a 9-node cluster, consisting of eight compute nodes and
one NVMeoF storage node. They are interconnected via a Mellanox FDR 40/56GbE
InfiniBand switch.  Each compute node is equipped with a single-port HCA, and
the storage node is equipped with two single-port HCAs. Each compute node has
two Intel Xeon Gold 5115 processors (20 vCPUs), 64~GB DRAM, SATA SSD for the OS
partition. The storage node has two Intel Xeon Silver 4215 processors (16
vCPUs), 128~GB DRAM, and an array of 24 Samsung PM9A3 U.2 960~GB NVMe SSDs. All
the nodes run Ubuntu 18.04.4, and the storage node runs PoseidonOS v0.12.0 for
NVMeoF management. We create up to 8 storage volumes and assign one to
each of the 8 compute nodes.

We evaluate the effect of offloading IO-intensive tasks with {\offloadFS} by
utilizing {\offloadDB} and {\offloadPrep}. For {\offloadDB}, we set the number
of client threads to 32, the number of background threads to 20, and the SSTable
size to 64~MB. For {\offloadPrep}, we set the number of pre-processing threads
to 4 for each initiator node. According to~\cite{mohan21vldb}, using 4 CPU cores
per GPU for pre-processing ensures that the GPU remains busy enough to avoid
data stalls for complex data models such as ResNet50.




For key-value store workloads, we run YCSB~\cite{cooper2010benchmarking}
benchmark. We load a 200~GB database using randomly generated 24-byte keys and
1~KB values as the initial state.  Thereafter, each workload of the query phase
submits 20 million requests in uniform distribution unless stated otherwise.
For ML pre-processing workloads, we run image classification workloads using
OpenImage 10~GB datasets~\cite{openimage}. {\offloadPrep} offloads the
pre-processing of a subset of a minibatch consisting of 64 images, and
temporarily stores the pre-processed results in memory for a short duration.
During an epoch, the training processes the entire dataset once and the
datasets used in ML training are generally larger than available memory.
Therefore, we disable the Offload Cache for {\offloadPrep}. 

\subsection{Comparison with Shared File Systems}\label{fs_eval-section}

\begin{figure}[!t]
\centering
\subfigure[{\offloadDB} with Write-only Workload]{
    \includegraphics[width=0.70\columnwidth]{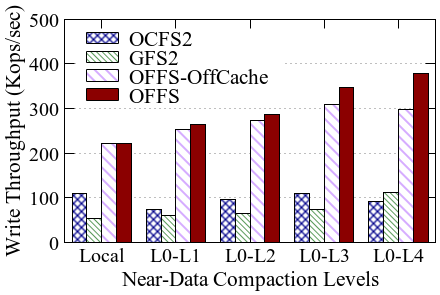}
	\label{fig:kv_load}
}
\subfigure[{\offloadPrep} with Read-only Workload]{
    \includegraphics[width=1.00\columnwidth]{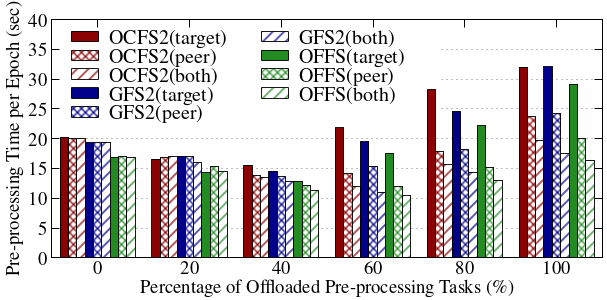}
	\label{fig:ai_load}
}
\caption{Comparison with Shared-Disk File Systems}
\label{fig:comparison}
\end{figure}

In the first set of experiments shown in Figures~\ref{fig:comparison}, we
compare the performance of {\offloadFS} against OCFS2 and GFS2. Since OCFS2 and
GFS2 are general-purpose shared-disk file systems, we implemented variants of
{\offloadDB} and {\offloadPrep} to make offloaded tasks perform IOs using POSIX
APIs.  
We measure the performance when pre-processing tasks are offloaded not only to
the storage node but also to a peer compute node (denoted as \texttt{peer}).
Additionally, we evaluate the performance of {\offloadPrep} when splitting the
offloaded tasks equally between a peer compute node and the storage node
(denoted as \texttt{both}). 


\subsubsection{{\offloadDB}}


In the experiment shown in Figure~\ref{fig:kv_load}, we create a single storage
volume from an array of 16 NVMe SSDs on PoseidonOS and measure the performance
of {\offloadDB} on a single compute node using YCSB Workload A with an adjusted
100\% write ratio in Zipfian distribution. 
We evaluate {\offloadFS} with and without using the Offload Cache on the target
node (denoted as \texttt{OFFS-OffCache} and \texttt{OFFS}).  When the Offload
Cache is not used, offloaded compaction tasks run slower on the storage node.  


When compaction tasks are not offloaded and all tasks are performed on the
single initiator node (denoted as \texttt{Local} on the X-axis), {\offloadFS} shows
approximately 2.2$\times$ higher throughput than OCFS2. This is because
{\offloadFS} is a lightweight user-level file system implemented with SPDK. 


When L0$\rightarrow$L1 compaction is offloaded (denoted as \texttt{L0-L1}),
OCFS2 exhibits a performance decrease.
This is due to the increased
overhead from concurrency control and coherence guarantees, as both the compute
and storage nodes are performing writes simultaneously.  OCFS2 controls
concurrent accesses to the same directory through directory locks, which
serializes offloaded tasks.  This approach simplifies lock management but
limits concurrency. OCFS2 shows slight performance improvements when offloading
more compaction tasks up to L3. However, 
notably, its performance is even
worse than the case where compaction tasks are not offloaded.

In contrast, GFS2 exhibits continuous performance improvements, though it
starts from a much lower baseline.  This is due to the way GFS2
manages concurrent IOs.  GFS2 manages concurrency at the file or block level,
allowing for more fine-grained control. Although this approach increases the
overhead of the distributed lock management and results in lower I/O
performance, GFS2 scales better than OCFS2 when multiple compute nodes
concurrently perform heavy writes.

When using {\offloadFS}, distributed locking  overhead is eliminated,
which results in higher throughput. Performance consistently improves as more
compactions are offloaded to the storage or peer node. By using
initiator-centric cache coherence, where concurrent access is simplified and controlled by the
application, {\offloadFS} achieves better
scalability than conventional shared-disk file systems.

{\offloadFS} performs best when offloading all compaction tasks to the peer
node, rather than the storage node.  However, OCFS2 and GFS2 perform better
with offloading to the storage node compared to offloading to peer node.  This
contrasting results are particularly interesting since it implies that
{\offloadFS} reduces the need for near-data processing.

When offloading to the peer compute node, network overhead increases because
file blocks are transmitted over the network. This also increases the network
latency of communication between daemon processes of OCFS2 and GFS2, which
eventually hurts the overall performance.  This underscores the need for
near-data processing because offloading tasks to the storage node eliminates
the need for transmitting file blocks over the network, which reduces the
communication latency between shared-disk file system daemon processes. In
contrast, {\offloadFS} minimizes file system metadata management and eliminates
delays associated with metadata-related messages.  Therefore, with
{\offloadFS}, it is more efficient to run tasks on nodes with higher
computational power, i.e., the benefits of near-data processing is not
significant.

In summary, OCFS2 performs best when compaction is not offloaded (110Kops with
\texttt{Local}).  GFS2 scales better than OCFS2, despite its higher overhead
and lower baseline performance.  {\offloadFS} provides consistently good
performance regardless of concurrency level, and performs best when all
compactions are offloaded (379Kops/sec with \texttt{L0-L4}), achieving a
3.36$\times$ performance improvement compared to the best performance of OCFS2.

\subsubsection{ML Pre-Processing}

In the experiments shown in Figure~\ref{fig:ai_load}, we vary the proportion of
images pre-processing tasks in each minibatch to be offloaded using
{\offloadPrep}. 
In this experiments, the storage node has a single
volume, and the peer compute node is completely idle, i.e., it does not perform
any task other than the offloaded ML pre-processing.

As both the compute and storage nodes run pre-processing tasks concurrently,
the pre-processing turn around time is determined by the slower node. When less
than half of the pre-processing tasks are offloaded, i.e., when the ratio of
offloaded tasks varies between 0 and 40\%, the preprocessing turn-around time
decreases, which is determined by the performance of the primary compute node,
where {\offloadPrep} is running.  
When more than 50\% of tasks are offloaded, the pre-processing time is affected
by the computational capability of offloadee nodes, i.e., the storage or peer
nodes. Unlike the {\offloadDB} results, ML preprocessing performs better when
offloaded to peer compute nodes rather than to storage nodes, even when using
OCFS2 or GFS2.  This is because ML preporcessing is more computation-intensive
than key-value stores. Additionally,  this is because the storage node runs the
PoseidonOS volume manager and has lower CPU specifications compared to the
compute node. 



However, it is noteworthy that using both the idle peer compute node and the
storage node together (denoted as \texttt{both}) results in better performance
compared to offloading only to the idle peer compute node.  This result
indicates that previous ML pre-processing studies~\cite{audibert23socc,
graur22cachew,mohan21vldb} can benefit from utilizing computing resources of
the storage node (and potentially computational SSDs and DPUs) as additional 
resources in addition to peer compute nodes.


We also note that the performance differences between the file systems
are less significant compared to {\offloadDB} results. Specifically, when using
only the target node, {\offloadFS} is approximately 1.85$times$ faster compared
to OCFS2 (15.19 sec vs. 28.18 sec).  This is because {\offloadPrep} only
performs read-only workloads. Therefore, the overhead of distributed lock
management in OCFS2 and GFS2 is minimal.

\subsection{{\offloadFS}: Scalability}\label{scale-section}

\begin{figure}[!t]
\centering
\subfigure[Throughput]{
    \includegraphics[width=0.46\columnwidth]{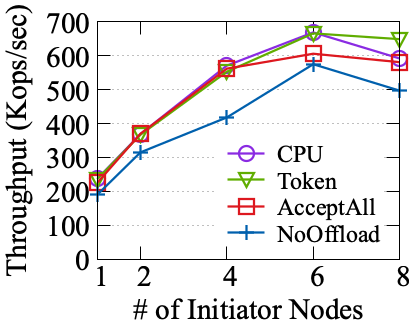}
	\label{fig:kv_thpt_scale}
}
\subfigure[CPU Utilization on Target]{
    \includegraphics[width=0.46\columnwidth]{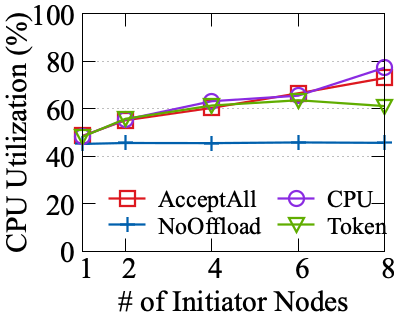}
	\label{fig:kv_cpu_scale}
}
\caption{
{\offloadDB} Scalability with YCSB A}
\label{fig:kv_scale}
\end{figure}

\begin{figure}[!t]
\centering
\subfigure[AI Epoch Time]{
    \includegraphics[width=0.46\columnwidth]{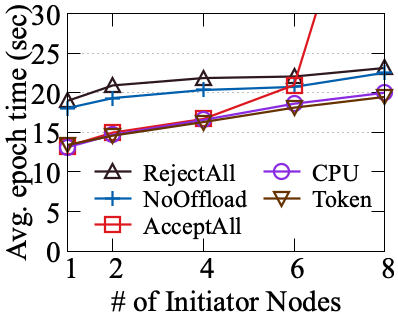}
	\label{fig:ai_time_scale}
}
\subfigure[CPU Utilization on Target]{
    \includegraphics[width=0.46\columnwidth]{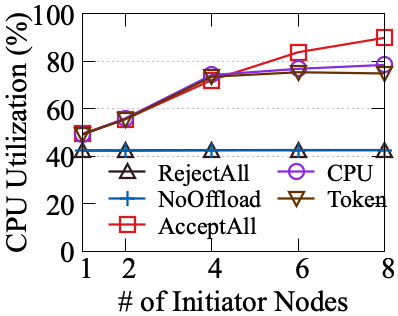}
	\label{fig:ai_cpu_scale_}
}
\caption{ML Pre-Processing Scalability
}
\label{fig:ai_scale}
\end{figure}

When a single storage node performs offloaded tasks for multiple {\offloadFS}
instances, it should selectively accommodate offload requests based on the
utilization of the storage node's computing resources.  In the experiments shown in
Figures~\ref{fig:kv_scale} and \ref{fig:ai_scale}, we vary the number of target
storage volumes and its corresponding initiator nodes from one to eight. 

\subsubsection{{\offloadDB}}

For {\offloadDB}, we run YCSB A (50\% Write in uniform distribution) workload on 
each initiator node, and L0$\rightarrow$L1 compaction is offloaded.
Overall, the total throughput increases as the number of {\offloadDB} instances
increases, up to 6 instances. However, when 8 instances are submitting heavy IO
requests concurrently to the storage node, performance drops due to
insufficient resources on the storage node. Even in the \texttt{NoOffload}
case, where compaction is not offloaded, performance degrades due to the
inherent limitations in PoseidonOS’s I/O throughput.  Compared to
\texttt{NoOffload}, \texttt{AcceptAll}, where all offload requests are
accepted, exhibits approximately twice higher the throughputs. It is noteworthy
that the CPU utilization averaged over the experiments on the storage node
increases no higher than 80\% even when eight {\offloadDB} instances offload
compactions simultaneously. 

The scheduling policy that uses tokens (denoted as \texttt{Token}) or the
\texttt{CPU} policy that rejects offload requests based on CPU load show about
10\% higher performance than \texttt{AcceptAll} when there are 6 {\offloadDB}
instances. In addition, when using tokens, performance degradation is minimal
even with 8 instances. On the other hand, the policy that rejects offload
requests based on CPU load exhibits performance drops due to the communication
latency caused by repeatedly requesting and rejecting offload requests.

\subsubsection{{\offloadPrep}}

Figure~\ref{fig:ai_scale} shows the scalability of offloading ML pre-processing
tasks. \texttt{NoOffload} shows the average epoch processing time when ML
pre-processing tasks are not offloaded but processed in local compute node.
When there is a single {\offloadPrep} instance running, each epoch takes about
18 seconds, but with 8 instances, the epoch time increases to approximately 22
seconds.
\texttt{AcceptAll} shows the performance when offload requests for 1/3
pre-processing tasks are sent to the storage node and all of them are accpted
and executed on the storage node, regardless of its resource usage. 
When 4 {\offloadPrep} instances share the storage node at the same time, 
\texttt{AcceptAll} is up to 27\% faster than \texttt{NoOffload}. 
However, when 8 {\offloadPrep} instances offload the tasks at the
same time, the execution time increases to about 60 seconds due to 
insufficient computing resources on the storage node.
\texttt{RejectAll} shows the performance when each initiator attempts to offload
1/3 of the pre-processing tasks to the storage node, but the storage node
rejects all the requests. This configuration aims to measure the performance
penalty incurred from sending offload requests that are rejected without any
benefit. Figure~\ref{fig:ai_cpu_scale_} shows that 
it is not significant even when all offload requests are rejected. 
\texttt{CPU} shows the performance when offload requests are allowed only if
CPU utilization is below 80\%. Otherwise, incoming requests are rejected.  In
the experiments, we observe that when 16 pre-processing RPC threads run
concurrently on the storage node, CPU usage exceeds 80\% and it leads to a
sharp decline in performance.
\texttt{Token} shows the performance when the storage node’s compaction engine
circulates  four tokens (each with a time-to-expire of 1 second) among the
initiators.  An initiator is permitted to submit offload requests only while it
holds a token.  \texttt{Token} shows comparable performance with \texttt{CPU},
However, the token policy performs about 3\% better than the CPU utilization
policy, which is due to fewer offload requests being rejected when using the
token policy.  However, the token-based policy requires parameter tuning to
determine how long a token remains valid, which is more challenging than tuning
based on CPU usage.



\subsection{{\offloadDB} Evaluation}

\subsubsection{Quantification of {\offloadDB} Design}

\begin{figure}[!t]
    \includegraphics[width=1\linewidth]{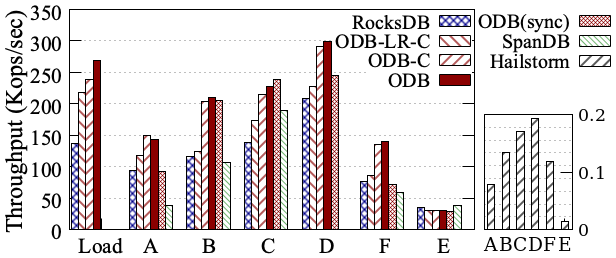}
    \caption{Quantifying Effect of OFfloadDB Designs and Performance Comparison}
    \label{fig:quant_throughput}
\end{figure}

In the experiments shown in Figure~\ref{fig:quant_throughput}, we run YCSB on a
single compute node with enabling each design of {\offloadDB} to quantify the
performance effect of each design.  We configure {\offloadDB} to offload
compaction tasks at all levels.  

\texttt{ODB-LR-C} represents the performance of {\offloadDB} with the Offload
Cache and Log Recycling disabled. If Log Recycling is disabled,
\texttt{ODB-LR-C} is essentially identical to RocksDB except that it offloads
compaction tasks, i.e., it shows how much performance gains can be achieved
solely by offloading compaction tasks to the storage node using {\offloadFS}.
In the Load workload, \texttt{ODB-LR-C} shows 1.51x higher throughput than
RocksDB as it reduces the write stall time.  \texttt{ODB-LR-C} is less
effective in the workload A, where 50\% of queries are reads.
For read-only workload C, \texttt{ODB-LR-C} still outperforms because SPDK lets
IO operations bypass the kernel. Additionally, compactions that were
scheduled in previous workloads are occasionally triggered during Workload C.
Therefore, \texttt{ODB-LR-C} consistently outperforms RocksDB for all
workloads. 

\texttt{ODB-C} represents the performance of {\offloadDB} with compaction
offloading and Log Recycling enabled, but the Offload Cache is disabled.  Log
Recycling reduces network usage by ensuring that the same key-value pair is
transmitted to the storage node only once.  Therefore, \texttt{ODB-C}
improves the write throughput by up to 9\% compared to \texttt{ODB-LR-C} in the
Load workload.  Log Recycling also contributes to the improvement in read
throughput of read-only workload C by approximately 40\%. This is due to the L0
cache on the initiator node. 

\texttt{ODB} represents the performance of {\offloadDB} with all designs
including the Offload Cache enabled. The Offload Cache improves the compaction
performance and improves the throughput of write-heavy workloads.
However, enabling the Offload Cache has little impact on read performance. 

YCSB workload E is a scan-intensive workload, and it is the only workload where
{\offloadDB} performs worse than the vanilla RocksDB. This is because
{\offloadFS} is not currently optimized for sequential scans. We leave
improving the scan performance of {\offloadDB} as future work.

\subsubsection{Comparative Performance}

In Figure~\ref{fig:quant_throughput}, we compare the performance of {\offloadDB}
with state-of-the-art \texttt{SpanDB}~\cite{Chen21spandb} and
\texttt{Hailstorm}~\cite{bindschaedler2020hailstorm}. 



\begin{figure}[!t]
\subfigure[Workload A]{
    \includegraphics[width=0.45\linewidth]{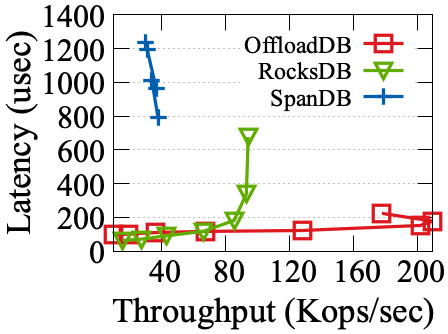}
	\label{fig:lt_ycsbA}
}
\subfigure[Workload C]{
	\includegraphics[width=0.45\linewidth]{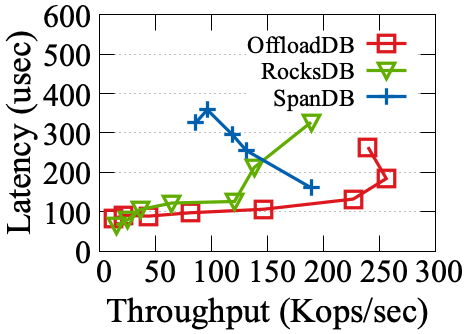}
	\label{fig:lt_ycsbC}
}
\caption{Latency-Throughput Analysis}
\label{fig:lt}
\end{figure}

\begin{figure*}[!t]
\centering
\begin{minipage}{0.8\linewidth}
    \subfigure[RocksDB]{
    \includegraphics[width=0.22\linewidth]{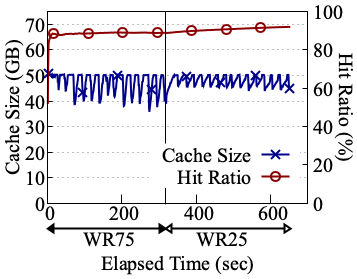}
    \label{fig:pollution_default}
    }
    \subfigure[dio-compaction]{
    \includegraphics[width=0.22\linewidth]{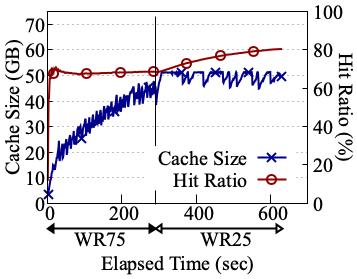}
    \label{fig:pollution_dio_comp}
    }
    \subfigure[{\offloadDB}]{
    \includegraphics[width=0.22\linewidth]{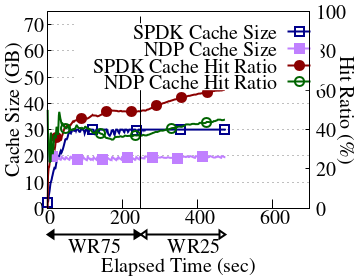}
    \label{fig:pollution_offload}
    }
    \subfigure[Throughput]{
    \includegraphics[width=0.22\linewidth]{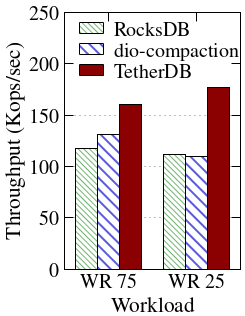}
    \label{fig:pollution_throughput}
    }
    \caption{Cache Usage and Hit Ratio over Time}
    \label{fig:mem_pollution}
\end{minipage}
\begin{minipage}{0.19\linewidth}
    \includegraphics[width=1\linewidth]{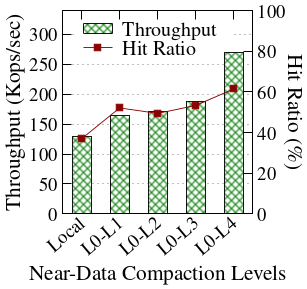}
    \caption{Impact of Offloading on Cache Pollution}
    \label{fig:pollution_impact}
\end{minipage}
\end{figure*}

SpanDB is a state-of-the-art key-value store that optimizes access to NVMe SSDs
using SPDK. It employs two storage tiers, i.e., small speed disks (SD) for the
WAL and lower levels of LSM trees, and capacity disks (CD).  For the speed disk,
we  use a local NVMe SSD (Samsung 970 Evo Plus). For the capacity disk (CD), we
use the disaggregated storage volume, formatted with EXT4, on the storage node.
Since SD is local, it yields lower latencies than CD.

Hailstorm is another state-of-the-art key-value store that offloads compactions
to remote peer compute nodes. Hailstorm runs on top of a distributed FUSE file
system with data striping. The data striping leads to high network usage due to
data transfers over the network. In contrast, {\offloadFS} is designed to reduce
network usage by enabling near-data processing in disaggregated storage. From
this perspective, the purpose of Hailstorm seems slightly differ from that of
{\offloadFS}. 



In our testbed, SpanDB is even outperformed by vanilla RocksDB for most
workloads. This is because SpanDB exclusively supports and operates in
failure-atomic {\it read-committed} mode, i.e., it flushes the WAL file every
time a new key-value pair is inserted.  In contrast, RocksDB and {\offloadDB}
call \texttt{fsync()} to flush dirty pages in a lazy manner by default because
not all applications require the immediate flush~\cite{Kim20Bolt,kim22listdb}.
Nonetheless, for a fair comparison, we configured {\offloadDB} to flush the WAL
file for each write. This configuration is denoted as \texttt{ODB(sync)}. Still,
SpanDB exhibits lower write throughputs than \texttt{ODB(sync)} primarily because
its parallel WAL logging and asynchronous request processing methods use a 
large number of foreground threads, which reduces the priority of background 
compaction tasks.


Hailstorm shows orders of magnitude lower performance than
other key-value stores (lower than 1Kops/sec throughputs for all workloads).
This is because its client-server communication, implemented using the Akka
toolkit, has concurrency issues, and its FUSE-based implementation incurs
multiple context switches per I/O operation. 




Figure~\ref{fig:lt} shows the latency-throughput curves of key-value stores.
RocksDB and {\offloadDB} follow the typical latency-throughput curve shape.
However, SpanDB shows abnormal curves because it gives priority to improving
throughput, utilizing asynchronous I/O, and dynamically adjusting the number of
background threads.

\subsubsection{Effect of OffloadCache}\label{isolated_caching-section}

In the experiments shown in Figures~\ref{fig:mem_pollution}, we run two YCSB
workloads sequentially. First, we execute a write-intensive \texttt{WR75}, a
variant of Workload A (25\%:75\% read-write ratio). Then, we run a
read-intensive \texttt{WR25}, another variant of Workload A (75\%:25\%
read-write ratio).


In the default configuration, both foreground clients and background compaction
tasks use buffered I/O, with the user-level block cache size set to the default
value of 8MB. Since both foreground and background IOs utilize the same page
cache, as shown in Figure~\ref{fig:pollution_default}, the workload starts with
the 50GB page cache filled with pages accessed by the previous Load workload.
Therefore, the cache hit ratio increases up to 92\% at the end of \texttt{WR25}
workload.  However, we note that about 67\% of those cache hits occur for
background compaction tasks.
While the write-intensive workload \texttt{WR75} is running, the page cache
usage frequently dips suddnely by up to 14GB, but quickly recovers.  This is
because victim SSTables are deleted from the page cache for each compaction.
But, soon the next compaction reads another set of victim SSTables and quickly
fills the page cache again. The pages from the victim SSTables pollute the
cache because they are not requested by foreground queries.

In the \texttt{dio-compaction} configuration, direct IO is used for compaction
tasks.  Therefore, the Load workload did not fill the page cache. As a result,
the page cache usage starts at approximately 3.3~GB, read queries gradually
fill up the page cache, and the cache hit ratio increases to about 80\%.  It is
noteworthy that the hit ratio is lower than the default configuration because
compaction tasks perform direct IOs. Nonetheless,
Figure~\ref{fig:pollution_throughput} shows that its throughput is not lower
than the default configuration.  This is because \texttt{dio-compaction} allows
to cache only hot blocks frequently accessed by foreground tasks and it helps
improve performance although direct IO may slow down background compaction
tasks. 

{\offloadDB} does not use the page cache but utilizes a user-level block cache
on the initiator node. Since it utilizes a Offload Cache on the storage node,
we configure {\offloadDB} to use less memory on the compute node than RocksDB.
Specifically, we set the L0 cache size, the block cache size, and the Offload
Cache size to 3~GB, 30~GB, and 20~GB, respectively.  Since the block cache size
is smaller, the cache hit ratio on the compute node (60\%) is lower than that
of \texttt{dio-compaction}.  However, due to L0 cache, the number of accesses
to SSTables is reduced by about 67.84\%, which results in a
throughput increase of 32.00\%, as shown in  
Figure~\ref{fig:pollution_throughput}.

In the experiments shown in Figure~\ref{fig:pollution_impact}, we measure the
block cache hit ratio and throughput of {\offloadDB} with varying the number of
offloaded compaction levels. We run YCSB Workload A (50\% Write) with the
Offload Cache disabled. If more compaction tasks are offloaded, cache
pollution problem becomes more serious when memory is scarce. So, we reduce the
block cache size of the compute node to 10GB for the experiments.  As we
offload more compaction tasks, more memory of the compute node is used for
foreground queries, and the cache hit rate increases accordingly.

\section{Conclusion}\label{conclusion-section}

This study identifies that NVMeoF-based disaggregated storage systems introduce
a new type of resource underutilization problem within storage nodes. To
address this challenge, we develop {\offloadFS}, a lightweight user-level file
system that allows compute nodes to offload tasks without the need for complex
cache coherence mechanisms.  In {\offloadFS}, the initiator node alone is
solely responsible for managing the file system's metadata, and the offloaded
tasks running on the target node have full access permissions to the blocks
authorized by the initiator. This {\it initiator-centric cache coherence}
greatly simplifies the resolution of I/O conflicts. Our performance study with
{\offloadDB} and ML pre-processing workloads shows that orchestraing different
types of IOs on initiator and target nodes separately helps leverage idle resources on storage nodes, thereby improving application performance.  {\it Upon publication of the paper, 
we will open-source 
the {\offloadFS} and {\offloadDB} source codes.
}

{\bibliographystyle{plain}
\bibliography{kvstores}}
\end{document}